\documentclass{pasa}%

\usepackage{graphicx}

\title[Modelling generalised Faraday rotation]{A phenomenological model for measuring generalised Faraday rotation}

\author[Lower, M.~E.]{Marcus~E.~Lower$^{1,2,}$\thanks{mlower@swin.edu.au}
\affil{$^1$Centre for Astrophysics and Supercomputing, Swinburne University of Technology, PO Box 218, Hawthorn, VIC 3122, Australia}%
\affil{$^2$Australia Telescope National Facility, CSIRO, Space and Astronomy, PO Box 76, Epping, NSW 1710, Australia}
}%

\jid{PASA}
\doi{10.1017/pas.\the\year.xxx}
\jyear{\the\year}

\usepackage{aas_macros}
\usepackage[
colorlinks=true,        
citecolor=blue,         
linkcolor=blue,         
urlcolor=blue           
]{hyperref}             

\begin{document}

\begin{frontmatter}
\maketitle

\begin{abstract}
Generalised Faraday rotation can induce frequency-dependent conversion between the linear and circular polarisation spectra of compact radio sources such as pulsars, fast radio bursts and active galactic nuclei.
I devise a simple phenomenological model that can be used to measure the effects of generalised Faraday rotation on the linearly and circularly polarised spectra of these sources.
The model is theory-agnostic, with an arbitrary wavelength dependence, and hence can accommodate for a variety of potential generalised Faraday rotation inducing media.
It can also be combined with additional observables to infer the physical properties of the intervening medium.
\end{abstract}

\end{frontmatter}

\section{Introduction}\label{sec:intro}

A generalised form of Faraday rotation occurs when the natural wave modes of the propagating medium are either linearly or elliptically polarised.
This process results in a conversion between the linear and circular polarisation components of the incident radiation \citep{Sazonov1969, Pacholczyk1970, Kennett1998}.
This phenomenon, referred to as Generalised Faraday rotation (GFR), has been observed in the radio spectra of active galactic nuclei (AGN) as a frequency-dependent variation between linear and circular polarisation \citep[e.g.][]{Macquart2002, OSullivan2013}.
GFR is also one of the suggested mechanisms responsible for correlated variations between the circular polarisation and phase-resolved rotation measures detected in some Galactic radio pulsars \citep{Noutsos2009, Dai2015, Ilie2019, Sobey2021}, and a handful of cosmological fast radio bursts \citep[FRBs;][]{Cho2020, Day2020}.

GFR can be induced from a variety of transmission processes.
This includes (but is not limited to) radiation propagating through relativistic plasma \citep{Kennett1998}, magnetic field reversals along the line of sight \citep{Melrose2010}, turbulence in AGN jets \citep{MacDonald2018} and vacuum birefringence induced by the strong magnetic fields of magnetars \citep{Heyl2002}.
Hence the detection of GFR can provide a novel means for probing the immediate environment surrounding Galactic neutron stars and FRB progenitors (see \citealt{Gruzinov2019} and \citealt{Vedantham2019} for discussions on using GFR to probe the local environment of FRB~20121102A).
However, unlike standard Faraday rotation, there is no widely used method for quantifying the impact of GFR on measured polarisation spectra. 

In this work, I introduce a phenomenological GFR model based around projecting the polarisation vector onto the Poincar\'{e} sphere.
I provide an overview of the implementation of this model in a preexisting software library and demonstrate its use in recovering an injected GFR signal in synthetic data.
I also discuss the limitations of the model, and how GFR measurements may be combined with other observables to infer the properties of the intervening medium.

\section{Modelling generalised Faraday rotation}

\begin{figure*}
    \centering
    \includegraphics[width=\linewidth]{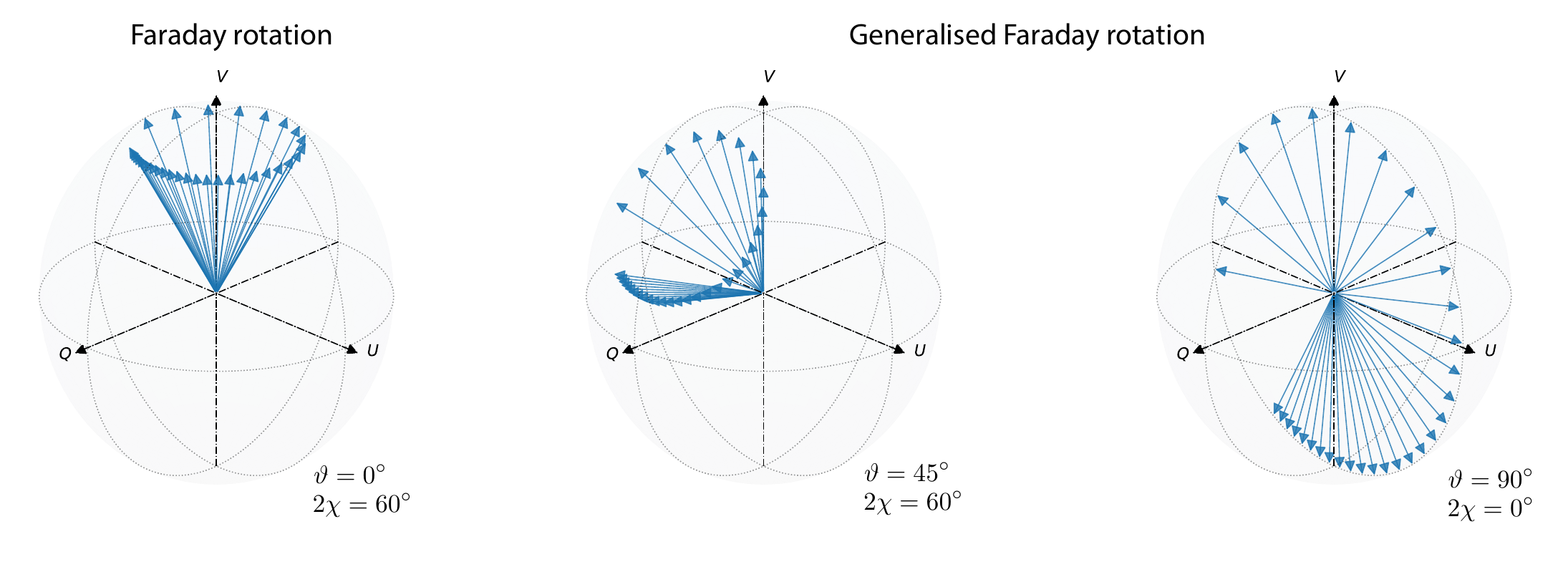}
    \caption{Polarization vector in terms of Stokes $Q$, $U$ and $V$ projected onto the Poincar\'{e} sphere, demonstrating the effects of standard Faraday rotation (left) and GFR (middle and right).}
    \label{fig:poincare}
\end{figure*}

The polarisation properties of electromagnetic radiation can be represented by four Stokes parameters, $\mathbf{S} = [I, Q, U, V]$, where $I$ is the total intensity, $Q$ and $U$ the two linear polarisations and $V$ is the circular polarisation. 
In general, Stokes $Q$, $U$ and $V$ can be described in spherical coordinates as 
\begin{equation}\label{eqn:stokes_params}
\begin{aligned}
    & Q = P\cos(2\Psi)\cos(2\chi),\\
    & U = P\sin(2\Psi)\cos(2\chi),\\
    & V = P\sin(2\chi),
\end{aligned}
\end{equation}
where $P = \sqrt{Q^2 + U^2 + V^2}$ is the total polarisation, $\Psi = \frac{1}{2}\tan^{-1}(U/Q)$ is the linear polarisation position angle (PA) and $\chi = \frac{1}{2}\tan^{-1}(V/\sqrt{Q^2 + U^2})$ is the ellipticity angle. 
It is often useful to project the Stokes $Q$, $U$ and $V$ components normalised by the total polarisation onto the Poincar\'{e} sphere.
Here, the polarisation vector, written as
\begin{equation}\label{eqn:pol_vec}
    \mathbf{P} = \frac{1}{P}
    \begin{bmatrix}
    Q\\
    U\\
    V
    \end{bmatrix}
    =
    \begin{bmatrix}
    \cos(2\Psi)\cos(2\chi)\\
    \sin(2\Psi)\cos(2\chi)\\
    \sin(2\chi)
    \end{bmatrix},
\end{equation}
is positioned on a sphere of unit radius with a co-latitude and co-longitude of $2\chi$ and $2\Psi$ respectively.

In standard Faraday rotation, the natural wave-modes of the medium through which the radiation has propagated are circularly polarised. 
This results in a wavelength dependent rotation of the polarisation vector about the V-axis at a fixed latitude on the Poincar\'{e} sphere, as depicted in the left-hand side of Figure~\ref{fig:poincare}.
Hence the polarisation vector in Equation~\ref{eqn:pol_vec} can be re-written as
\begin{equation}\label{eqn:poln_wave}
    \mathbf{P}(\lambda) = 
    \begin{bmatrix}
    \cos[2\Psi(\lambda)]\cos(2\chi)\\
    \sin[2\Psi(\lambda)]\cos(2\chi)\\
    \sin(2\chi)
    \end{bmatrix},
\end{equation}
where the wavelength dependence of the PA is given by
\begin{equation}\label{eqn:rm}
    \Psi(\lambda) = \Psi_{0} + {\rm RM} (\lambda^{2} - \lambda_{c}^{2}).
\end{equation}
Here, $\Psi_{0}$ is the PA at some reference wavelength $\lambda_{c}$ (often chosen to be the centre frequency of the input spectrum) and ${\rm RM}$ is the rotation measure which is dependent on the line-of-sight electron column density ($n_{e}$) and parallel magnetic field strength ($B_{\parallel}$) as
\begin{equation}
    {\rm RM} = \frac{e^{2}}{8 \pi^{2} \varepsilon_{0} m_{e}^{2} c^{3}} \int d \ell n_{e} B_{\parallel},
\end{equation}
where $e$ is the electron charge, $\varepsilon$ is the vacuum permittivity, $m_{e}$ the electron rest mass, $c$ the vacuum speed of light, $\ell$ the path length.

\begin{figure*}
    \centering
    \includegraphics[width=0.63\linewidth]{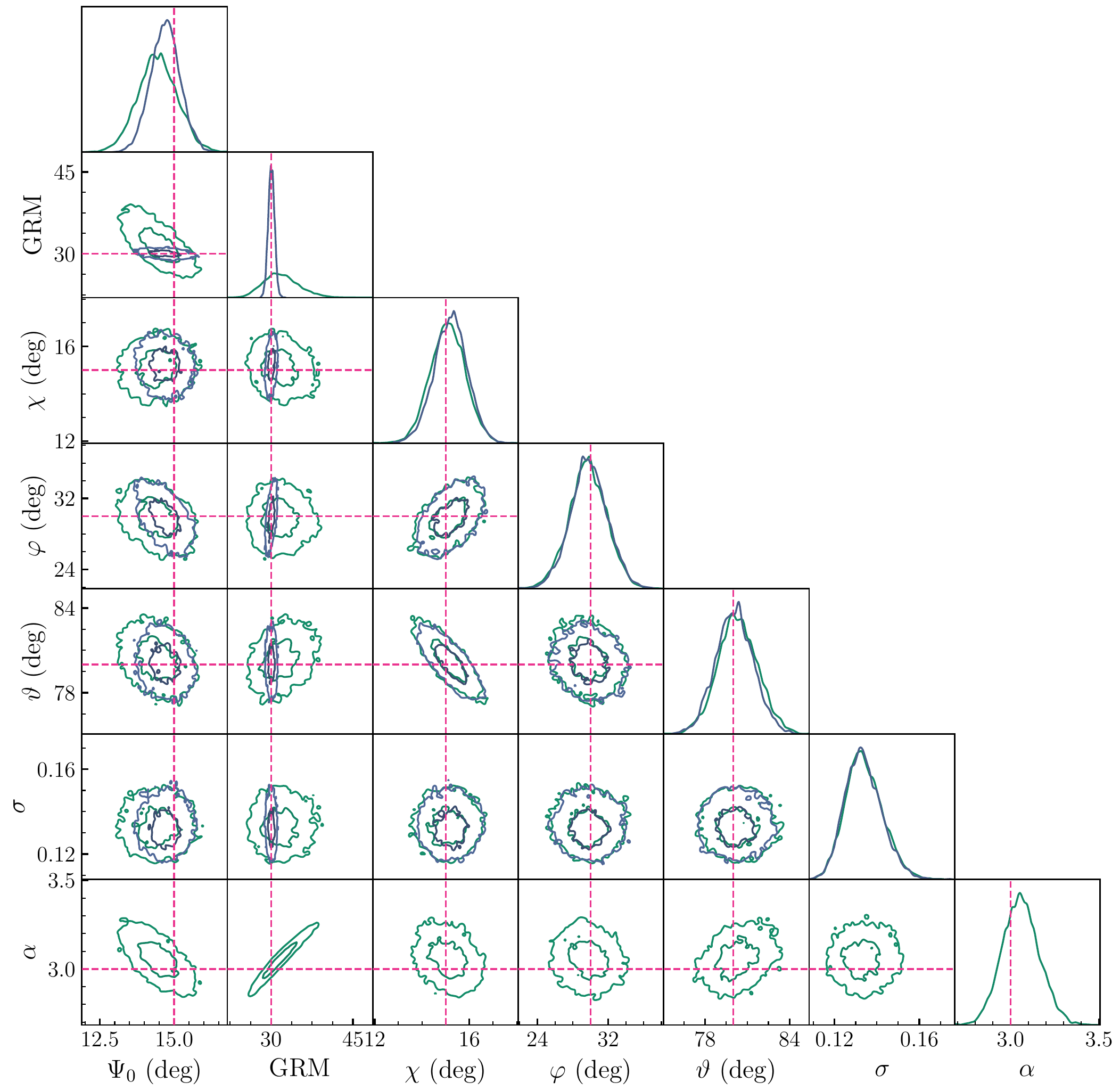}
    \caption{Recovered posterior distributions after from fitting the simulated data with $\alpha$ fixed to three (blue) and $\alpha$ sampled as a free parameter (green). Dashed magenta lines indicate the injected values.}
    \label{fig:corner}
\end{figure*}

Under GFR, the natural wave modes of the medium are either elliptically or linearly polarised, resulting in a rotation of the polarisation vector about an arbitrary point on the Poincar\'{e} sphere. 
Examples of this effect are shown in the middle and right-hand side of Figure~\ref{fig:poincare}.
This tilting of the polarisation plane can be emulated through the addition of two rotation matrices
\begin{equation}
    \mathbf{R}_{\vartheta} = 
    \begin{bmatrix}
    \cos(\vartheta) & 0 & \sin(\vartheta)\\
    0 & 1 & 0\\
    -\sin(\vartheta) & 0 & \cos(\vartheta)
    \end{bmatrix},
\end{equation}
and
\begin{equation}
    \mathbf{R}_{\varphi} = 
    \begin{bmatrix}
    \cos(\varphi) & -\sin(\varphi) & 0\\
    \sin(\varphi) & \cos(\varphi) & 0\\
    0 & 0 & 1
    \end{bmatrix},
\end{equation}
where the angles $\vartheta$ and $\varphi$ represent respective rotations about the Stokes $U$ and $V$ axes.
Hence, the full phenomenological GFR model can be written as
\begin{equation}\label{eqn:pol_model}
    \hat{\mathbf{P}}_{m}(\lambda)
    =
    \mathbf{R}_{\vartheta} \cdot \mathbf{R}_{\varphi} \cdot \mathbf{P}(\lambda).
\end{equation} 
The wavelength dependence of $\Psi$ in GFR can take on different values depending on the underlying physics that governs the propagating medium. 
Hence, Equation~\ref{eqn:rm} is re-written with an arbitrary wavelength exponent, $\alpha$, such that
\begin{equation}\label{eqn:grm}
    \Psi(\lambda) = \Psi_{0} + {\rm GRM} (\lambda^{\alpha} - \lambda_{c}^{\alpha}),
\end{equation}
where ${\rm GRM}$ is the generalised rotation measure with units of rad\,m$^{-\alpha}$.

\section{Parameter estimation}

To measure the impact of GFR on some observed polarisation data, the GFR-model can be fit directly to the Stokes $Q$, $U$ and $V$ spectra.
Posterior probability distributions for the model parameters can be inferred from using Bayes' theorem as
\begin{equation}
    p(\theta | d) = \frac{{\cal L}(d | \theta) \pi(\theta)}{{\cal Z}(d)},
\end{equation}
where $d$ is the input Stokes spectra, $\theta$ the model parameters, ${\cal L}(d | \theta)$ is the likelihood function, $\pi(\theta)$ are our priors on the model parameters and ${\cal Z}(d)$ is the Bayesian evidence.
A list of the model parameters and their associated priors are given in Table~\ref{tab:priors}.
\begin{table}[t]
    \centering
    \caption{Standard priors on the model parameters}
    \label{tab:priors}
    \begin{tabular}{lcc}
    \textrm{Parameter}&
    \textrm{Prior type}&
    \textrm{Range (units)}\\
    \hline
    $\alpha$ & Uniform & $0$--$10$\\
    GRM & Uniform & $0$--$1000$ (rad\, m$^{-\alpha}$)\\
    $\Psi_{0}$ & Uniform & $-90$--$90$ (deg)\\
    $\chi$ & Uniform & $0$--$45$ (deg)\\
    $\vartheta$ & Uniform & $0$--$180$ (deg)\\
    $\varphi$ & Uniform & $-180$--$180$ (deg)\\
    $\sigma$ & Uniform & $0$-$1$\\
    \hline
    \end{tabular}
\end{table}
An implementation of the GFR model is available for use via the \texttt{rmnest} package~\citep{Bannister2019, Lower2020}\footnote{\href{https://github.com/mlower/rmnest}{https://github.com/mlower/rmnest}}, where the input data is fit using a Gaussian likelihood of the form 
\begin{equation}
    {\cal L}(\mathbf{P}(\lambda) | \theta) = \prod_{i}^{N} \frac{1}{\sqrt{2\pi\sigma^{2}}} \exp \Big[ -\frac{(\mathbf{P}(\lambda_{i}) - \hat{\mathbf{P}}_{m}(\theta; \lambda_{i}))^{2}}{2\sigma^{2}} \Big].
\end{equation}
Here, $\mathbf{P}(\lambda)$ is the input polarisation spectra, $\hat{\mathbf{P}}_{m}(\lambda; \theta)$ is the GFR-model from Equation~\ref{eqn:pol_model} and $\sigma$ is either the off-pulse root-mean-square of the fractional polarisation or a free parameter that approximates the variance of the data.
The posterior distributions are sampled using the Bayesian inference library, \texttt{Bilby}~\citep{Ashton2019}, with the \texttt{PyMultiNest} nested sampling algorithm~\citep{Buchner2014}.
This implementation has been used to successfully model the GFR detected in an unusual repeat burst from FRB~20201124A (Kumar et al. in prep.).

\begin{figure}
    \centering
    \includegraphics[width=\linewidth]{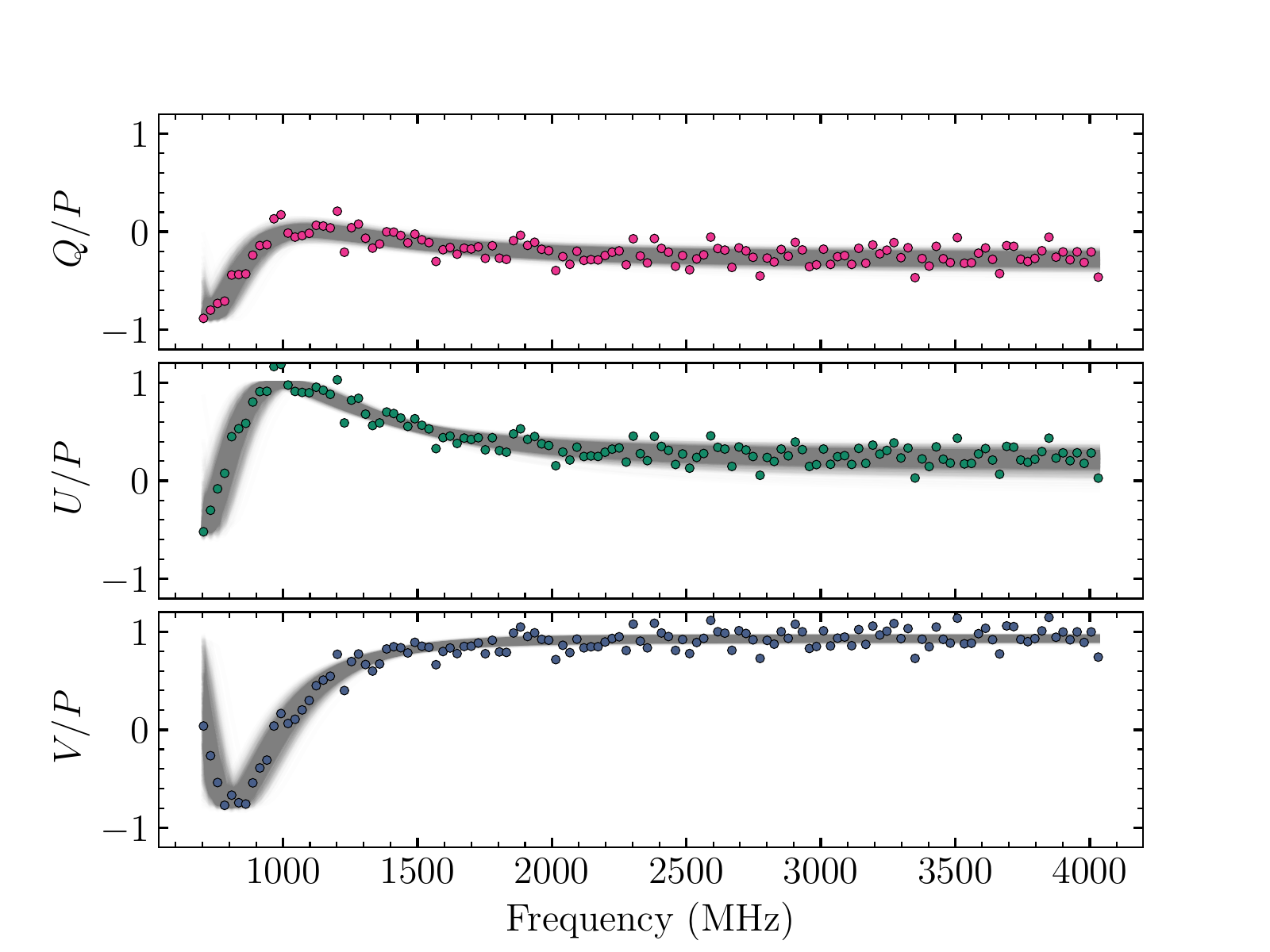}
    \caption{Injected Stokes spectra (points) and 1000 random draws (traces) from the posteriors of the GFR-model where $\alpha$ was treated as a free parameter.}
    \label{fig:fit}
\end{figure}

To demonstrate the models use, I injected a synthetic GFR spectrum into Gaussian noise covering a frequency bandpass of 704-4032\,MHz to emulate a broadband polarisation spectrum one might detect with the Parkes Ultra-Wideband Low (UWL) receiver system \citep{Hobbs2020}. 
Figure \ref{fig:corner} presents the resulting posterior distributions after fitting both a GFR model in which $\alpha$ was fixed at three (the fixed-$\alpha$ model) and a variant where $\alpha$ was included as a free parameter (the free-$\alpha$ model). 
The injected values are well recovered within the 68\,percent credible intervals. 
Figure \ref{fig:fit} shows the simulated Stokes $Q$, $U$ and $V$ spectra along with random draws from the free-$\alpha$ model posteriors, where it is clear the recovered GFR model matches the injected signal.

\section{Discussion}

\subsection{Limiting factors}

There is a strong covariance between GRM and $\alpha$ in the posteriors for the free-$\alpha$ model. 
The result of this covariance can be seen in Figure~\ref{fig:corner} as the broader, correlated posterior distribution for the GRM when compared to the fixed-$\alpha$ model. 
This behaviour is expected, as from Equation \ref{eqn:grm} a large GRM with a small $\alpha$ could be partially confused for a large $\alpha$ and small GRM.
Hence for polarisation spectra obtained from relatively instruments with low fractional bandwidths or data capture systems with limited frequency resolution, GFR fits with a free-$\alpha$ may be incapable of constraining the values of $\alpha$ and the GRM with high precision.
Limited observing bandwidths may also result in any GFR-induced circular polarisation being indistinguishable from circular polarisation that is intrinsic to the source, e.g. from overlapping orthogonal polarisation modes with different spectral indices \citep[e.g.][]{Ramachandran2004}.
Similar issues may arise if the intrinsic emission from the source is band-limited, a phenomena that appears to be a characteristic of bursts from repeating FRB sources \citep{Pleunis2021}.

Astrophysical sources of linearly polarised radiation are often affected by `standard' Faraday rotation as their emitted radio waves propagate through the magnetised interstellar medium of the Galaxy. In the case of extra-galactic sources, additional Faraday rotation can arise from the host environment, inter-galactic medium and circum-galactic medium of any intervening galaxies. 
The GFR-fitting framework described here relies upon any normal Faraday rotation having already been accounted for and removed from the polarisation spectra being analysed. 
Failure to do so would result in either a non-detection or biased measurements of any GFR that is present in the data. 
Additionally, strong GFR can also result in an overestimation of the RM associated with standard Faraday rotation. 
This could be mitigated through a simultaneous, joint-fit to both the standard and generalised Faraday rotation and by inspecting the projection of the polarisarion spectra on the Poincar\'{e} sphere ahead of performing a GFR-fit to the Faraday de-rotated data.

\subsection{Properties of the intervening medium}

Unlike standard Faraday rotation, where the RM can be related back to the physical properties of the relatively cold, magnetised intervening medium, there is no `standard' relation for the generalised case.
However, measurements of the GRM and $\alpha$ can be used to constrain the type intervening medium responsible for the GFR in the data.
For example, a relativistic plasma is expected to result in a rotation of the polarisation plane away from the Stokes $V$ pole along with a $\lambda^{3}$ dependence \citep{Kennett1998}.
The angle $\vartheta$ provides a means to infer the ellipticity of the natural wave modes in the GFR-inducing medium. 
For $\vartheta = 0^{\circ}$ variations in the polarisation vector are restricted to the $Q$-$U$ plane, implying circularly polarised natural modes (i.e standard Faraday rotation if $\alpha = 2$), while linearly polarised natural modes would return $\vartheta = 90^{\circ}$. 
Conflating the detection and measurement of GFR with other observables can enable us to obtain much stronger constraints on on the physical properties of the intervening medium.
For instance, excess dispersion caused by an increase in the line-of-sight electron column density can be used to infer the density of a GFR-inducing plasma, as well as the perpendicular magnetic field strength along the line of sight (see section 3.3 of \citealt{Li2019}).

\section{Conclusion}

In this work I have described a phenomenological method for modelling the effects of generalised Faraday rotation in the polarised spectra of pulsars and FRBs. 
The method defined here will be useful in exploring the spectra of these objects, particularly given the recent (and up-coming) commissioning of broadband radio receiver systems on large radio telescopes such as the Parkes UWL \citep{Hobbs2020} and telescope arrays with large fractional bandwidths \citep[e.g. MeerKAT;][]{Bailes2020}.
Observations with such instruments may allow for us to infer whether any detected circular polarisation is induced by GFR or is intrinsic to their emission mechanism.
As the model is theory-independent it can be used to model the effects of GFR induced from a wide variety of intervening media without the need to adapt complex polarised radiative transfer equations for use in parameter estimation.
Its simplicity means it can also be easily applied to polarisation spectra obtained from any astrophysical or terrestrial source of electromagnetic radiation.

\section*{Acknowledgments}
This work was supported by the Australian Research Council (ARC) Laureate Fellowship FL150100148 and the ARC Centre of Excellence CE170100004 (OzGrav).
M.E.L. receives support from the Australian Government Research Training Program and CSIRO Space and Astronomy.
I thank R.~M.~Shannon and S.~Johnston for their helpful suggestions.

\bibliographystyle{pasa-mnras}
\bibliography{paper.bib}

\end{document}